\documentclass[aps,pre,preprint,groupedaddress]{revtex4-1}
\usepackage{amsmath,amssymb,xcolor,graphicx}

\begin{document}

\title{Hamiltonian fluid reductions of drift-kinetic equations and the correspondence with water-bag distribution functions}

\author{M. Perin}
\author{C. Chandre}
\author{E. Tassi}

\affiliation{Aix-Marseille Universit\'e, Universit\'e de Toulon, CNRS, CPT UMR 7332, 13288 Marseille, France}

\date{\today}

\begin{abstract}
Hamiltonian models for the first three moments of the drift-kinetic distribution function, namely the density, the fluid velocity and the parallel pressure, are derived from the Hamiltonian structure of the drift-kinetic equations. The link with the water-bag closure is established, showing that, unlike the one-dimensional Vlasov equations, these solutions are the only Hamiltonian fluid reductions for the drift-kinetic equation. These models are discussed through their equations of motion and their Casimir invariants.
\end{abstract}


\maketitle

\section{Introduction}

An effective model for describing the dynamics of strongly magnetized plasmas is provided by the drift-kinetic equations (see, e.g. Ref.~\cite{Haz03}). These describe the evolution of distribution functions of the guiding centers of the plasma particles and apply to phenomena where the characteristic frequencies are much lower than the cyclotron frequency and where the characteristic wave lengths are much longer than the Larmor radius of the particle species under consideration. The advantage of the drift-kinetic description with respect to the more complete Vlasov description, is that in the former, the distribution functions, at the leading order, are defined on a reduced phase space, when compared to the latter. Indeed for drift-kinetic distribution functions the dependence on the coordinate corresponding to the gyration angle of the particles around the magnetic field, is ignored, because its evolution takes place on time scales too short for the phenomena of interest.

In this article we consider a simple drift-kinetic system in slab geometry. We assume that the evolution of a particle species (ions or electrons) in the plasma is described by the drift-kinetic equation
\begin{equation}
\label{dkorig}
\frac{\partial \tilde{f}}{\partial t}=-v\frac{\partial \tilde{f}}{\partial z}+\frac{q}{m}\frac{\partial \phi}{\partial z}\frac{\partial \tilde{f}}{\partial v}-\frac{c}{B}[\phi , \tilde{f}]+\frac{\mu c }{q l}[x,\tilde{f}].
\end{equation}
Such equation can be obtained, for instance, from the gyrokinetic equation of Ref.~\cite{S_2010}, in the limit of small Larmor radii.

In Eq.~\eqref{dkorig}, we indicate with $\tilde{f}(\mathbf{r},v,\mu,t)$ the guiding center distribution function of the particle species under consideration, where $\mathbf{r}=(x,y,z)$ denotes the position in configuration space, $v$ corresponds to the velocity of the particles along the direction of the magnetic field, $\mu$ is the particle magnetic moment and $t$ denotes time. The magnetic field is assumed to be straight and weakly inhomogeneous in space. Its expression is  given by $\mathbf{B}(x,y)=B(1 - x/l)\hat{z}$, where $B$ is a constant and $l$ is a characteristic length of variation of the magnetic field, such that $x/l \ll 1$.  As the magnetic field is along $z$, we will refer to $z$ as the longitudinal direction and $x$ and $y$ as the transverse directions. The constants $c$, $q$ and $m$ indicate the speed of light, the particle charge and mass, respectively. The canonical Poisson bracket $[\cdot,\cdot]$ is defined by
\begin{equation*}
[g,h]=\frac{\partial g}{\partial x}\frac{\partial h}{\partial y}-\frac{\partial g}{\partial y}\frac{\partial h}{\partial x},
\end{equation*}
for all functions $g$ and $h$. We also assume that the distribution function $\tilde{f}$ be of the form
\begin{equation*}
\tilde{f}(\mathbf{r},v,\mu,t)=m f(\mathbf{r},v,t)\delta(\mu-\mu_0)/(2 \pi B),
\end{equation*}
that is we assume a beam-like distribution with respect to the magnetic moment. Making use of this assumption, from Eq.~\eqref{dkorig} one easily obtains the following evolution equation for the function $f(\mathbf{r},v,t)$:

\begin{equation}
\label{eqDK}
\frac{\partial f}{\partial t}=-v\frac{\partial f}{\partial z}+\mathsf{q}\frac{\partial \phi}{\partial z}\frac{\partial f}{\partial v}-[\phi ,f]+\mathsf{q}\frac{\partial f}{\partial y},
\end{equation}
where the value of the physical constants $c, m ,B ,\mu_0$ and $l$ is set equal to unity and where, to keep track of the particle charge, we introduced the symbol $\mathsf{q}$, whose value corresponds to $1$ or $-1$ depending on whether the drift-kinetic species consists of ions or electrons, respectively. Equation~\eqref{eqDK} is complemented by an equation relating the electrostatic potential $\phi$ to $f$. The results presented in the paper hold for a generic equation of the type $\phi=\mathcal{L}\int \mathrm{d}v f$, where $\mathcal{L}$ is some self-adjoint linear operator. This allows the treatment of different specific examples, just by choosing different expressions for $\mathcal{L}$. For instance, the case of a Poisson's equation for drift-kinetic electrons with a neutralizing ion background corresponds to
\begin{equation*}
\mathcal{L}\int f\ \mathrm{d}v =\Delta^{-1}\left(\int f\ \mathrm{d}v-1\right).
\end{equation*}
On the other hand, the case of a quasi-neutrality relation with drift-kinetic ions and adiabatic electrons, as given in Ref.~\cite{Morel07}, corresponds to
\begin{equation*}
\mathcal{L}\int f\ \mathrm{d}v =\int f\ \mathrm{d}v-1.
\end{equation*}
With regard to the physical interpretation of the terms appearing in Eq.~\eqref{eqDK}, we remark that the first three terms correspond to the one-dimensional Vlasov-Poisson equation which describes the dynamics of the distribution function along $z$ and $v$. In particular, the second and third term account for the advection and for the action of the electric force, respectively, along the longitudinal direction. The last two terms, on the other hand, account for the $E\times B$ and the $\nabla B$ drifts respectively, which affect the evolution of the distribution function in the direction transverse to the strong magnetic field. 

A way of solving the drift-kinetic equation is to use fluid reductions, that describe the time evolution of $P_i(\mathbf{r},t)$, namely the $i$-th fluid moment of the distribution function with respect to the velocity $v$, such that
\begin{equation}
\label{eqP}
P_i(\mathbf{r},t)=\int v^if(\mathbf{r},v,t)\ \mathrm{d}v.
\end{equation}
The use of fluid reductions of kinetic models is ubiquitous in plasma physics. Indeed, whereas the drift-kinetic equation involves the distribution function of particles $f$ defined in the four-dimensional phase space $(\mathbf{r},v)$, fluid models rely on quantities such as the density $\rho=P_0$, the velocity $u=P_1/P_0$ and the pressure $p=P_2-P_1^2/P_0$ which are defined solely in the three-dimensional configuration space $\mathbf{r}$. In addition of being more tangible than their kinetic counterpart, fluid models require less numerical resources to be solved. On the other hand, fluid models derived from kinetic equations lead to an infinite set of moment equations. Specific closures have to be performed in order to truncate this infinite hierarchy of moments resulting from the reduction procedure. A way of closing the system is to compute the closure in order to recover some kinetic effects such as the growth rate of kinetic instabilities (see e.g. Refs.~\cite{Hammett90} and \cite{Sarazin09}). Our purpose in this article is to find closures that preserve the geometrical features of the drift-kinetic equation. Indeed, Eq.~\eqref{eqDK} can be cast into a Hamiltonian form. Not preserving this Hamiltonian structure during the fluid reduction may lead to models that exhibit some non-physical dissipation. A Hamiltonian fluid model for the first two moments of the distribution function derived from the drift-kinetic equation has been found in Ref.~\cite{Tassi14}. Indeed, if we consider, e.g., functionals of the first two fluid moments, namely functionals of the kind $F[P_0,P_1]$, we have to perform a closure on $P_2$ by expressing it as a function of $P_0$ and $P_1$ such that $P_2=P_2(P_0,P_1)$. Moreover, it has been shown that in order to preserve the Hamiltonian structure of the parent model, namely the drift-kinetic equation, one has to impose $P_2=P_1^2/P_0+\mathcal{A}P_0^3$, where $\mathcal{A}$ is a constant\footnote{It is worth noting that the complete closure is actually $P_2=P_1^2/P_0+\mathcal{A}P_0^3+\mathcal{B}$. However, the case $\mathcal{B}\neq0$ is not physically relevant. Indeed, introducing the pressure $p=P_2-P_1^2/P_0=\mathcal{A}P_0^3+\mathcal{B}$, one can see that $p\rightarrow\mathcal{B}$ as $P_0\rightarrow0$. As $P_0$ corresponds to the density of electrons, $\mathcal{B}$ is the pressure of the void and as a consequence we shall only consider the case $\mathcal{B}=0$.}. This closure corresponds to a single water-bag distribution function of height $\mathcal{A}$, namely a distribution function of the type
\begin{multline*}
f(\mathbf{r},v,t)=\mathcal{A}\bigg(\Theta\left[v-\frac{P_1(\mathbf{r},t)}{P_0(\mathbf{r},t)}+\frac{P_0(\mathbf{r},t)}{12\mathcal{A}}\right]\\
-\Theta\left[v-\frac{P_1(\mathbf{r},t)}{P_0(\mathbf{r},t)}-\frac{P_0(\mathbf{r},t)}{12\mathcal{A}}\right]\bigg),
\end{multline*}
where $\Theta$ is the Heaviside distribution (see e.g. Ref.~\cite{Besse08}). A closure preserving the Hamiltonian structure for an arbitrary number of moments in the "$\delta f$" approximation of the drift-kinetic equation has been presented in Ref.~\cite{Tassi15}. So far it has not been generalized to the fully non-linear case.

In this article, we derive a fluid model for the first three moments of the fully non-linear drift-kinetic distribution function. We consider the density $\rho$, the velocity $u$ and the second reduced moment $S_2$ as dynamical variables (see Ref.~\cite{Perin15}). Compared to the two field model, the three field model has the advantage of including $S_2$ (or equivalently $P_2$) as a dynamical variable. This is significant since the Hamiltonian depends explicitly on $P_2$ through the kinetic energy. We construct a closure on the third and fourth moments respectively such that the resulting fluid model preserves the Hamiltonian structure of the parent kinetic one. Furthermore, we investigate the role of the dimensionality of the system in the closure procedure. Indeed, the drift-kinetic model, which accounts for the evolution of a distribution function defined in a four-dimensional phase space, represents a useful step between the one-dimensional Vlasov-Amp\`ere equations (see Ref.~\cite{Perin14}) and the full Vlasov-Maxwell system. The method developed hereby makes use of some Mathematica\copyright{} code which is provided in the appendix in order for the reader to be able to reproduce the calculations.

\section{Hamiltonian fluid reduction of the drift-kinetic equation}

The drift-kinetic model described by Eq.~\eqref{eqDK} can be cast in a Hamiltonian form by introducing the bracket given in Ref.~\cite{Tassi14}
\begin{equation}
\label{eqBrackDK}
\{F,G\}=\int f\left(\frac{\partial F_f}{\partial z}\frac{\partial G_f}{\partial v}-\frac{\partial F_f}{\partial v}\frac{\partial G_f}{\partial z}-\mathsf{q}[F_f,G_f]\right)\ \mathrm{d}\mathbf{r}\mathrm{d}v,
\end{equation}
where $F_f=\delta F/\delta f$ denotes the functional derivative of $F$ with respect to $f$ such that
\begin{equation*}
F[f+\delta f]-F[f]=\int\frac{\delta F}{\delta f}\delta f\ \mathrm{d}\mathbf{r}\mathrm{d}v+\mathcal{O}\left(\delta f^2\right).
\end{equation*}
The Hamiltonian writes
\begin{equation}
\label{eqHamDK}
\mathcal{H}[f]=\frac{1}{2}\int f\left(v^2+2[1-x]+\mathsf{q}\mathcal{L}\int f\ \mathrm{d}v'\right)\ \mathrm{d}\mathbf{r}\mathrm{d}v,
\end{equation}
in which the first term accounts for the kinetic energy of the particles in the longitudinal direction while the other terms account for the kinetic energy of the particles in the direction perpendicular to the magnetic field. Equation~\eqref{eqDK} is then obtained by using $\partial_tf=\{f,\mathcal{H}\}$.

Using Eq.~\eqref{eqP} to perform the change of variables from the distribution function $f$ to the fluid moments $P_i$, Bracket~\eqref{eqBrackDK} and Hamiltonian~\eqref{eqHamDK} write respectively
\begin{multline}
\label{eqBrackP}
\{F,G\}=\sum\limits_{n,m\in\mathbb{N}}\int\bigg(nP_{n+m-1}\left(G_n\frac{\partial F_m}{\partial z}-F_n\frac{\partial G_m}{\partial z}\right)\\
-\mathsf{q}P_{n+m}[F_m,G_n]\bigg)\ \mathrm{d}\mathbf{r},
\end{multline}
and
\begin{equation}
\label{eqHamP}
\mathcal{H}[P_0,P_1,P_2]=\frac{1}{2}\int \left(P_2+P_0\left[2(1-x)+\mathsf{q}\mathcal{L}P_0\right]\right)\ \mathrm{d}\mathbf{r},
\end{equation}
where $F_m=\delta F/\delta P_m$ denotes the functional derivative of $F$ with respect to $P_m$. Bracket~\eqref{eqBrackP} can be rewritten by introducing the operator $\mathcal{J}$ such that
\begin{equation}
\label{eqBrackP2}
\{F,G\}=\sum\limits_{n,m\in\mathbb{N}}\int F_n\mathcal{J}_{nm}G_m\ \mathrm{d}\mathbf{r},
\end{equation}
where $\mathcal{J}_{nm}=-m\partial_zP_{n+m-1}-(n+m)P_{n+m-1}\partial_z+\mathsf{q}[P_{n+m},\cdot]$.

The fluid reduction process described above allows us to replace the distribution function of particles $f$ defined in phase space with an infinite number of fluid moments $P_i$ defined in configuration space. In order to be able to compute the dynamics of these variables, it is mandatory to truncate the infinite hierarchy of moments. Due to the terms $P_{n+m-1}$ and $P_{n+m}$ in Bracket~\eqref{eqBrackP2}, functionals of the kind $F[P_0,\dots,P_N]$ for any $N\geq 1$ do not constitute a Poisson subalgebra of the bracket. In other words, the bracket between two functionals of the type $F[P_0,\dots,P_N]$ not only depends on $(P_0,\dots,P_N)$ but also on $N$ higher order moments $(P_{N+1},\dots,P_{2N})$. As a consequence, one has to express these higher order moments $P_{N+1\leq n\leq 2N}$ with respect to the variables $P_{n\leq N}$ in order to be able to close the infinite hierarchy of moments.

When considering functionals of the type $F[P_0,P_1,P_2]$, Bracket~\eqref{eqBrackP2} becomes
\begin{equation}
\label{eqBrackP3}
\{F,G\}=\sum\limits_{n,m=0}^2\int F_n\mathcal{J}_{nm}G_m\ \mathrm{d}\mathbf{r}.
\end{equation}
Denoting $F_\chi=(F_0,F_1,F_2)$ and defining the inner product
\begin{equation*}
\langle\mathbf{a},\mathbf{b}\rangle=\sum\limits_{n=0}^2\int a_nb_n\ \mathrm{d}\mathbf{r},
\end{equation*}
Bracket~\eqref{eqBrackP3} reads $\{F,G\}=\langle F_\chi,\mathcal{J}G_\chi\rangle$, and can be split into a longitudinal and a transverse (with respect to the magnetic field) part such that $\{F,G\}=\{F,G\}_\parallel+\{F,G\}_\perp$, where  $\{F,G\}_\parallel=\langle F_\chi,\mathcal{J}_\parallel G_\chi\rangle$ with
\begin{equation*}
\mathcal{J}_\parallel=-
\begin{pmatrix}
0 & & \partial_z P_0 & & 2\partial_zP_1\\
0 & & \partial_zP_1 & & 2\partial_zP_2\\
0 & & \partial_zP_2 & & 2\partial_zP_3
\end{pmatrix}
-
\begin{pmatrix}
0 & & P_0 & & 2P_1\\
P_0 & & 2P_1 & & 3P_2\\
2P_1 & & 3P_2 & & 4P_3
\end{pmatrix}
\partial_z,
\end{equation*}
corresponds to the one-dimensional Vlasov bracket found in Refs.~\citep{Gibbons81,Gibbons08} that accounts for the dynamics along the direction of the magnetic field, namely $z$. The bracket corresponding to the transverse part reads $\{F,G\}_\perp=\langle F_\chi,\mathcal{J}_\perp G_\chi\rangle$, where
\begin{equation*}
\mathcal{J}_\perp=\mathsf{q}
\begin{pmatrix}
[P_0,\cdot] & [P_1,\cdot] & [P_2,\cdot]\\
[P_1,\cdot] & [P_2,\cdot] & [P_3,\cdot]\\
[P_2,\cdot] & [P_3,\cdot] & [P_4,\cdot]
\end{pmatrix}
,
\end{equation*}
and describes the dynamics in the transverse direction to the magnetic field, namely the $xy$-plane. Like in the one-dimensional Vlasov case, $\mathcal{J}_\parallel$ introduces a contribution of the third order moment $P_3$ which does not belong to the subset of functional of the form $F[P_0,P_1,P_2]$. In addition, $\mathcal{J}_\perp$ introduces a contribution of the fourth order moment $P_4$ which constitutes an important difference with the one-dimensional Vlasov equation. In this case, the closure has to be performed on both $P_3$ and $P_4$ as functions of $P_0$, $P_1$ and $P_2$ such that $P_3=P_3(P_0,P_1,P_2)$ and $P_4=P_4(P_0,P_1,P_2)$. We want to compute these closures such that the resulting fluid model is Hamiltonian, i.e., in particular that Bracket~\eqref{eqBrackP3} satisfies the Jacobi identity, namely
\begin{equation*}
J=\{F,\{G,H\}\}+\{H,\{F,G\}\}+\{G,\{H,F\}\}=0,
\end{equation*}
for any functionals $F$, $G$ and $H$. $J=0$ provides the necessary and sufficient condition on $P_3$ and $P_4$ for the closure to be Hamiltonian. $J$ can be decomposed in the following way as
\begin{multline*}
J=\{F,\{G,H\}_\parallel\}_\parallel+\{F,\{G,H\}_\perp\}_\parallel\\
+\{F,\{G,H\}_\parallel\}_\perp+\{F,\{G,H\}_\perp\}_\perp+\circlearrowleft,
\end{multline*}
where $\circlearrowleft$ means that the summation is performed over all the circular permutations of the functionals $F$, $G$ and $H$. The two brackets $\{\cdot,\cdot\}_\parallel$ and $\{\cdot,\cdot\}_\perp$ have to satisfy the Jacobi identity independently from each other, i.e., we must have $J_\parallel=\{F,\{G,H\}_\parallel\}_\parallel+\circlearrowleft=0$ and $J_\perp=\{F,\{G,H\}_\perp\}_\perp+\circlearrowleft=0$. Indeed, $\mathcal{J}_\parallel$ and $\mathcal{J}_\perp$ are both bilinear differential operators of order one. However, while $\mathcal{J}_\parallel$ differentiates along the $z$ direction, $\mathcal{J}_\perp$ only involves derivatives in the transverse, respectively $x$ and $y$, directions. As these directions are independent from one another, it is required that both brackets satisfy the Jacobi identity. Furthermore, it has been shown in Ref.~\cite{Perin14} that in order for the bracket $\{\cdot,\cdot\}_\parallel$ to satisfy the Jacobi identity (or equivalently to make $J_\parallel$ vanish), we must have
\begin{equation}
\label{eqP3}
P_3=3\frac{P_1P_2}{P_0}-2\frac{P_1^3}{P_0^2}+P_0^4S_3\left(\frac{P_2}{P_0^3}-\frac{P_1^2}{P_0^4}\right),
\end{equation} 
where $S_3$ is an arbitrary function of its argument. This result suggests the introduction of the reduced moments, defined in Ref.~\cite{Perin15} as
\begin{align*}
\rho&=\int f\ \mathrm{d}v,\\
u&=\frac{1}{\rho}\int vf\ \mathrm{d}v,\\
S_2&=\frac{1}{\rho^3}\int (v-u)^2f\ \mathrm{d}v.
\end{align*}
The change from the variable $\chi=(P_0,P_1,P_2)$ to $\psi=(\rho,u,S_2)$ is given by
\begin{equation*}
\rho=P_0,\qquad u=\frac{P_1}{P_0}, \qquad S_2=\frac{P_2}{P_0^3}-\frac{P_1^2}{P_0^4}.
\end{equation*}
The inverse of this transformation is given by
\begin{equation*}
P_0=\rho,\qquad P_1=\rho u, \qquad  P_2=\rho u^2+\rho^3S_2.
\end{equation*}
By using the chain rule, noting that
\begin{align*}
\frac{\delta F}{\delta P_0}&=\frac{\delta F}{\delta \rho}-\frac{u}{\rho}\frac{\delta F}{\delta u}+\left(\frac{u^2}{\rho^3}-3\frac{S_2}{\rho}\right)\frac{\delta F}{\delta S_2}\\
\frac{\delta F}{\delta P_1}&=\frac{1}{\rho}\frac{\delta F}{\delta u}-2\frac{u}{\rho^3}\frac{\delta F}{\delta S_2}\\
\frac{\delta F}{\delta P_2}&=\frac{1}{\rho^3}\frac{\delta F}{\delta S_2},
\end{align*}
Bracket~\eqref{eqBrackP3} becomes
\begin{equation}
\label{eqBrackS}
\{F,G\}=\langle F_\psi,\mathcal{K}G_\psi\rangle,
\end{equation}
where $\mathcal{K}$ can be decomposed such that $\mathcal{K}=\mathcal{K}_\parallel+\mathcal{K}_\perp$ and where
\begin{multline*}
\mathcal{K}_\parallel=
\begin{pmatrix}
0 & & 0 & & 0\\
0 & & 0 & & \rho^{-1}\partial_zS_2\\
0 & & -\rho^{-1}\partial_zS_2 & & -2\partial_z\left(\rho^{-2}S_3\right)
\end{pmatrix}
\\
-
\begin{pmatrix}
0 & & 1 & & 0\\
1 & & 0 & & 0\\
0 & & 0 & & 4\rho^{-2}S_3(S_2)
\end{pmatrix}
\partial_z,
\end{multline*}
is associated with the bracket in the longitudinal direction and
\begin{widetext}
\begin{multline*}
\mathcal{K}_\perp=-\mathsf{q}
\begin{pmatrix}
0 & & 0 & & 0\\
0 & & \left[S_2,\rho\right] & & 3\rho^{-1}\left[S_3,\rho\right]+\rho^{-1}[u,S_2]\\
0 & & \rho^{-1}[S_3,\rho]-\rho^{-1}[u,S_2] & & 6[S_2^2,\rho^{-1}]-3\left[S_4,\rho^{-1}\right]-2\left[u,\rho^{-2}S_3\right]
\end{pmatrix}
\\
+\mathsf{q}
\begin{pmatrix}
[\rho,\cdot] & & [u,\cdot] & & [S_2,\cdot]\\
[u,\cdot] & & \rho^{-2}\left[\rho^3S_2,\cdot\right] & & \rho^{-4}\left[\rho^4S_3,\cdot\right]\\
[S_2,\cdot] & & \rho^{-4}[\rho^4S_3,\cdot] & & \rho^{-6}\left[\rho^5S_4,\cdot\right]+4\rho^{-2}S_3[u,\cdot]-3\rho^{-4}[\rho^3S_2^2,\cdot]
\end{pmatrix}
,
\end{multline*}
\end{widetext}
leads to the definition of the bracket in the transverse direction. The Hamiltonian of the system writes
\begin{equation}
\label{eqHamS}
\mathcal{H}[\rho,u,S_2]=\frac{1}{2}\int \left(\rho u^2+\rho^3S_2+\rho\left[2(1-x)+\mathsf{q}\mathcal{L}\rho\right]\right)\ \mathrm{d}\mathbf{r}.
\end{equation}
Expressed in terms of the variables, namely $(\rho,u,S_2)$, the closure on $P_3$ given by Eq.~\eqref{eqP3} eventually writes $S_3(S_2)$. The generic closure on $P_4$ is now replaced by the corresponding reduced moment $S_4$ which reads
\begin{equation*}
S_4=\frac{P_4(\rho,u,S_2)}{\rho^5}-\frac{u^4}{\rho^4}-6\frac{u^2}{\rho^2}S_2-4\frac{u}{\rho}S_3(S_2)=S_4(\rho,u,S_2).
\end{equation*}

As stated previously, in order for Bracket~\eqref{eqBrackS} to satisfy the Jacobi identity we must have $J_\parallel=J_\perp=\{F,\{G,H\}_\perp\}_\parallel+\{F,\{G,H\}_\parallel\}_\perp+\circlearrowleft=0$ where $\{F,G\}_\perp=\langle F_\psi,\mathcal{K}_\perp G_\psi\rangle$ and $\{F,G\}_\parallel=\langle F_\psi,\mathcal{K}_\parallel G_\psi\rangle$. Due to the constraint $S_3=S_3(S_2)$, we already have $J_\parallel=0$. We now look for constraints in order to make $J_\perp=\{F,\{G,H\}_\perp\}_\perp+\circlearrowleft$ vanish. This is done by computing the Jacobi identity for particular choices of the functionals $F$, $G$ and $H$. The calculations are lengthy and, consequently, are done with Mathematica\cite{Mathematica}. In Appendix, we provide the details that allow us to compute the necessary and sufficient constraints on $S_3$ and $S_4$ in order for Bracket~\eqref{eqBrackS} to satisfy the Jacobi identity. These constraints write respectively
\begin{equation}
\label{eqS3}
3S_2-(S_3')^2+4S_3S_3''=0,
\end{equation}
and
\begin{equation}
\label{eqS4}
S_4=\frac{1}{5}\left(4S_3S_3'+9S_2^2\right)+\frac{\mathcal{C}}{\rho^5},
\end{equation}
where $\mathcal{C}$ is a constant. Equations~\eqref{eqS3} and \eqref{eqS4} are necessary and sufficient conditions in order to cancel $J_\perp$. In addition, it is shown in Appendix that they are sufficient conditions in order for Bracket~\eqref{eqBrackS} to satisfy the Jacobi identity. We notice that the fluid closures for the drift-kinetic equation are more constrained than in the one-dimensional Vlasov equation where any function $S_3=S_3(S_2)$ leads to a Hamiltonian closure.

\section{Correspondence with water-bags}

As shown in the previous section, Hamiltonian models with three moments derived from the drift-kinetic equation are those whose third and fourth order reduced moments, namely $S_3$ and $S_4$, satisfy Eqs.~\eqref{eqS3} and \eqref{eqS4} respectively. However, in order for $S_4$, which corresponds to the kurtosis of the distribution function, to stay bounded as the particle density $\rho$ tends to 0, one has to impose $\mathcal{C}=0$ in Eq.~\eqref{eqS4}.

\begin{figure}
\centering
\includegraphics[width=0.5\textwidth]{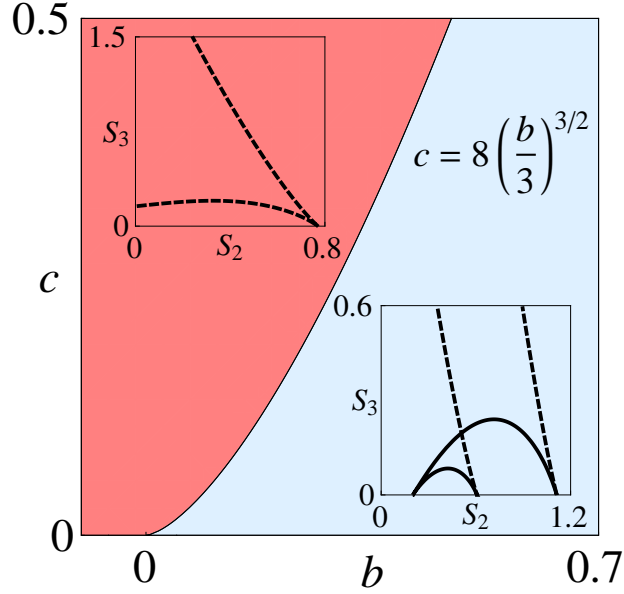}
\caption{\label{figBC}Diagram representing the nature of the parametric solutions given by Eqs.~\eqref{eqS2Para} and \eqref{eqS3Para} as a function of $b$ and $c$. In each regions, the insets represent the solutions $S_3(S_2)$. All the quantities are expressed in arbitrary units.}
\end{figure}

Assuming that there exists an open set in which the solution of Eq.~\eqref{eqS3} is of constant sign, it can be transformed into an Emden-Fowler equation (see e.g. Ref.~\cite{Polyanin03}) whose solution is given in parametric form
\begin{align}
\label{eqS2Para}
S_2(\lambda)&=\frac{4}{3}b^2+c\lambda+\frac{b}{2}\lambda^2-\frac{1}{64}\lambda^4,\\
\label{eqS3Para}
S_3(\lambda)&=\pm\left(c+b\lambda-\frac{1}{16}\lambda^3\right)^2,
\end{align}
where $\lambda$ parametrizes the curve $(S_2(\lambda),S_3(\lambda))$ and where $b$ and $c$ are constant. This implies that the solution $S_3(S_2)$ is given by parts such that $S_3$ is of constant sign. $S_3(\lambda)$ has either one or three real roots. The solutions $S_3(S_2)$ to Eq.~\eqref{eqS3} can be sorted out accordingly depending on the sign of $\Delta=64b^3-27c^2$. Moreover, the roots of $S_3(\lambda)$ are such that $S_2\geq0$, which is a necessary condition as $S_2$ corresponds to the variance of the distribution function. Indeed, let $\lambda_0$ be such that $S_3(\lambda_0)=0$. We thus have
\begin{equation*}
S_2(\lambda_0)=\frac{3}{64}\left(\lambda_0^2-\frac{16}{3}b\right)^2\geq0.
\end{equation*}
The case $S_2(\lambda_0)=0$ corresponds to $\Delta=0$, otherwise $S_2(\lambda_0)>0$. For $\Delta<0$, the solution $S_3(S_2)$ to Eq.~\eqref{eqS3} has two branches with a single root. For $\Delta>0$, it has four branches and three roots. This is summarized in Fig.~\ref{figBC} which represents the nature of the solutions depending on the position in the parameter space $(b,c)$.
It is worth noting that $\Delta$ is even in $c$ such that Fig.~\ref{figBC} is symmetric for $c<0$ through the change of parameter $\lambda\rightarrow-\lambda$. Moreover, for all $b<0$, we have $\Delta<0$ such that the corresponding solutions have only one real root (thus belong to the dark pink region in Fig.~\ref{figBC}).

Each inset in Fig.~\ref{figBC} exhibits two kinds of solutions. The branches represented by dashed lines are such that $S_3\neq0$ for $S_2=0$. From the definition of the second order reduced moment $S_2\propto\int(v-u)^2f\ \mathrm{d}v$, which corresponds to the variance of the distribution function, we see that $S_2=0$ implies $f=0$. Consequently, this requires that $S_3$, which corresponds to the skewness of the distribution function, vanishes. Hence, branches of solutions represented in dashed lines are not physically relevant. On the other hand, the solutions represented by solid lines and which are physically relevant correspond to water-bag closures as we show below. We recall that the water-bag distribution function for three moments corresponds to a function of the type
\begin{multline*}
f(\mathbf{r},v,t)=a_1\left[\Theta\left(v-\mathrm{v}_1(\mathbf{r},t)\right)-\Theta\left(v-\mathrm{v}_2(\mathbf{r},t)\right)\right]\\
+(a_1+a_2)\left[\Theta\left(v-\mathrm{v}_2(\mathbf{r},t)\right)-\Theta\left(v-\mathrm{v}_3(\mathbf{r},t)\right)\right],
\end{multline*}
where $a_1$ and $(a_1+a_2)$ are the heights of the two water-bags respectively and $\mathrm{v}_1(\mathbf{r},t)$, $\mathrm{v}_2(\mathbf{r},t)$ and $\mathrm{v}_3(\mathbf{r},t)$ are the contours of the bags (see e.g. Refs.~\citep{Morel07,Besse08,Gravier08}). The associated closure $S_3^{WB}=S_3^{WB}(S_2^{WB})$ is given in Ref.~\cite{Perin15b} in parametric form, namely
\begin{align*}
S_2^{WB}(n_1)&=\frac{a_1^2+6a_2a_1n_1^2+4a_2(a_2-a_1)n_1^3-3a_2^2n_1^4}{12a_1^2(a_1+a_2)^2},\\
S_3^{WB}(n_1)&=-\frac{a_2(n_1-1)^2n_1^2(a_1+a_2n_1)^2}{4a_1^3(a_1+a_2)^3},
\end{align*}
where $n_1\in[0;1]$ is the density of particles in the first bag which parametrizes the curve $(S_2^{WB}(n_1),S_3^{WB}(n_1))$. One can verify that for any value of the parameters $b$ and $c$ within the light blue region in Fig.~\ref{figBC} (i.e. the region where $\Delta>0$) there exists a change of parameters that leads to the water-bag closure. Indeed, this is done by performing the invertible change $\lambda\mapsto\alpha n_1+\beta$ and solving the equations $S_2(\alpha n_1+\beta)=S_2^{WB}(n_1)$ and $S_3(\alpha n_1+\beta)=S_3^{WB}(n_1)$ which leads to
\begin{align*}
b&=\frac{a_1^2 + a_1 a_2 + a_2^2}{12 a_1 |a_2|(a_1 + a_2)},\\
c&=\frac{(a_2-a_1)(2a_1+a_2)(2a_2+a_1)}{54 [a_1 |a_2| (a_1 + a_2)]^{3/2}},\\
\lambda&=\frac{2 (a_1 - a_2)+6 a_2n_1}{3 \sqrt{a_1 |a_2| (a_1 + a_2)}}.
\end{align*}

It is shown, e.g., in Ref.~\cite{Morel14} that the water-bag distribution function is preserved by the dynamics given by the drift-kinetic equation. Hence, any fluid model derived from it is Hamiltonian for an arbitrary number of fields. Furthermore, as it has been shown, the water-bag closure corresponds to the only physically relevant closure with three moments derived from the drift-kinetic equation given by Eq.~\eqref{eqDK} as it is the case with two moments. Whether this result extends to higher numbers of moments is an open question.

\begin{figure}
\centering
\includegraphics[width=0.5\textwidth]{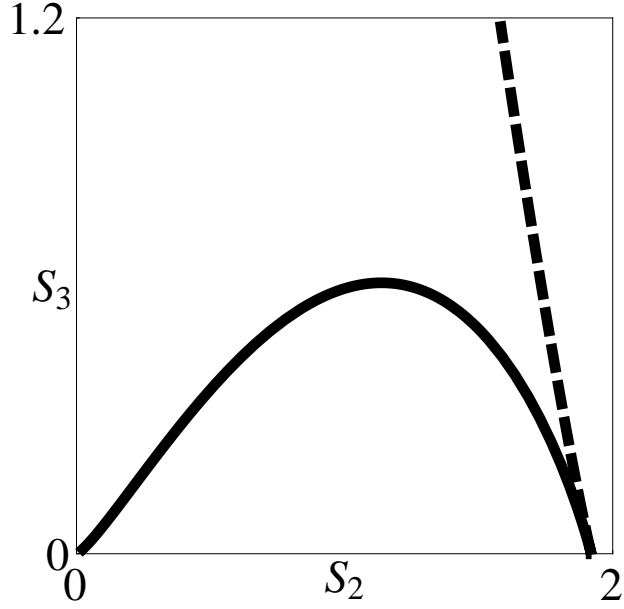}
\caption{\label{figLimit}$S_3$ (given by Eq.~\eqref{eqS3Para}) as a function of $S_2$ (given by Eq.~\eqref{eqS2Para}) for $b=0.4$ and $c=8(b/3)^{3/2}$ (which corresponds to $\Delta=0$). All the quantities are expressed in arbitrary units.}
\end{figure}

For $\Delta=0$, there is a one-parameter family of solutions which do not exactly correspond to the water-bags. More precisely, these solutions, which are given by Eqs.~\eqref{eqS2Para} and \eqref{eqS3Para} with $c=8(b/3)^{3/2}$, constitute the limiting case $a_2/a_1\rightarrow+\infty$ and the corresponding closures are represented on Fig.~\ref{figLimit}.
These solutions have two branches, one physically irrelevant depicted in dashed line and one physically relevant for which $S_3=0$ when $S_2=0$.

\section{The Hamiltonian three moment drift-kinetic model}

The bracket and the Hamiltonian of the three moment model are given respectively by Eq.~\eqref{eqBrackS} and Eq.~\eqref{eqHamS}. The corresponding equations of motion are then obtained by using $\dot{F}=\{F,\mathcal{H}\}$ which leads to
\begin{align*}
\frac{\partial\rho}{\partial t}&=-\frac{\partial(\rho u)}{\partial z}-[\mathcal{L}\rho,\rho]+\mathsf{q}\frac{\rho}{y},\\
\frac{\partial u}{\partial t}&=-u\frac{\partial u}{\partial z}-\frac{1}{\rho}\frac{\partial(\rho^3S_2)}{\partial z}-\mathsf{q}\frac{\partial(\mathcal{L}\rho)}{\partial z}-[\mathcal{L}\rho,u]+\mathsf{q}\frac{u}{y},\\
\frac{\partial S_2}{\partial t}&=-u\frac{\partial S_2}{\partial z}-\frac{1}{\rho^3}\frac{\partial(\rho^4S_3)}{\partial z}-[\mathcal{L}\rho,S_2]+\mathsf{q}\frac{S_2}{y}.
\end{align*}
It is more common to consider the pressure $p=\rho^3S_2$ as dynamical variable instead of $S_2$. The corresponding equation of motion writes
\begin{equation*}
\frac{\partial p}{\partial t}=-u\frac{\partial p}{\partial z}-3p\frac{\partial u}{\partial z}-\frac{\partial(\rho^4S_3)}{\partial z}-[\mathcal{L}\rho,p]+\mathsf{q}\frac{p}{y}.
\end{equation*}
The equations of motion are very much like the ones for the one-dimensional Vlasov equation. The only difference comes from the advection term associated with the transverse drifts (two rightmost terms in the above equations). As a consequence, if one just adds these terms in the equations of motion while keeping the closure used with the one-dimensional fluid model resulting from the Vlasov equation (see Ref.~\cite{Perin14}), it is very much likely that the Hamiltonian structure is lost as the drift-kinetic closure is much more constrained. This would result in a dissipative system. The consequence of this can be seen by considering the Casimir invariants as we show below.

Casimir invariants $C$ of a Poisson bracket $\{\cdot,\cdot\}$ are quantities that commute with all functionals $F$, namely $\{F,C\}=0$ for all $F$. In the three field model, one can show that the system possesses three Casimir invariants given by
\begin{align*}
C_1&=\int\rho\ \mathrm{d}\mathbf{r},\\
C_2&=\int\rho\kappa_0\ \mathrm{d}\mathbf{r},\\
C_3&=\int\left(u-\frac{1}{4}\rho\kappa_0^2\right)\ \mathrm{d}\mathbf{r},
\end{align*}
where $\kappa_0'(S_2)=1/\sqrt{2S_3}$. $C_1$ and $C_2$ correspond respectively to the total mass and the total entropy of the system. $C_3$ is referred to as the generalized velocity. If the Hamiltonian structure is lost, $C_2$ and $C_3$ are no longer conserved in general which is at the origin of the fake dissipation introduced in the model. For water-bags, these Casimir invariants can be expressed as functions of the water-bags dynamical variables $\mathrm{v}_i$. Indeed, they are linear combinations of $\int\mathrm{v}_1\ \mathrm{d}\mathbf{r}$, $\int\mathrm{v}_2\ \mathrm{d}\mathbf{r}$ and $\int\mathrm{v}_3\ \mathrm{d}\mathbf{r}$ which are the Casimir invariants of the two water-bag model.

One can show that if $S_3\neq0$ (the case $S_3=0$ is trivial as it is not Hamiltonian), the previous quantities satisfy $\{F,C\}=0$ for all $F$ if and only if Eqs.~\eqref{eqS3} and \eqref{eqS4} are satisfied or equivalently if and only if the associated bracket satisfies the Jacobi identity. Therefore, there is an equivalence between the model being Hamiltonian and the fact that it possesses as many Casimir invariants as dynamical field variables. This feature is strongly linked to the fact that the possible Hamiltonian closures are the water-bag ones. In other words, if advection terms associated with transverse drifts are added to the equations of motion without ensuring that the closure used is Hamiltonian (or equivalently water-bag), two conserved quantities are lost and non-physical diffusion across surfaces defined by these quantities occur.

\section{Summary}

In summary, starting from the drift-kinetic equation, we derive a Hamiltonian model for the first three moments of the distribution function, namely the density, the fluid velocity and the pressure. The computation of the Jacobi identity provides constraints on the fluid closures in order for the resulting model to be Hamiltonian. The solutions of these equations are given in parametric form, which allow us to identify them with the water-bag closure which, in this case, corresponds to the only Hamiltonian closures. Even though the resulting equations of motion differ only by the terms associated with the transverse drifts from those derived from the one-dimensional Vlasov equation, we show that the closure is more constrained in the drift-kinetic case.

\begin{acknowledgments}
This work was supported by the Agence Nationale de la Recherche (ANR GYPSI) and the French Research Federation for fusion study.
\end{acknowledgments}

\appendix*
\section{Computation of $J$}

The computation of the constraints given by Eqs.~\eqref{eqS3} and \eqref{eqS4}, which are necessary and sufficient conditions in order for Bracket~\eqref{eqBrackS} to satisfy the Jacobi identity, is done by using Mathematica\cite{Mathematica}. In this appendix, we provide the detailed code along with the procedure required in order to obtain these constraints.

We need to include the "VariationalMethods" package in order to be able to compute functional derivatives. This is done with the command-line
\begin{verbatim}
Needs["VariationalMethods`"];
\end{verbatim}
We then define the coordinate we are considering, which in this article correspond to $x$, $y$ and $z$. This is done with the command-line
\begin{verbatim}
r={x,y,z};
\end{verbatim}
We define then the field variables of the system, $\psi=(\rho,u,S_2)$:
\begin{verbatim}
psi={rho[x,y,z],u[x,y,z],Subscript[S,2][x,y,z]};
\end{verbatim}
Then, we proceed to the definition of a set of three functional derivatives, respectively $F_\psi$, $G_\psi$ and $H_\psi$ which we, for sake of simplicity, denote \verb|dF|, \verb|dG| and \verb|dH| respectively.
\begin{verbatim}
dF={Subscript[F,rho][x,y,z],Subscript[F,u][x,y,z],
  Subscript[F,2][x,y,z]};
dG={Subscript[G,rho][x,y,z],Subscript[G,u][x,y,z],
  Subscript[G,2][x,y,z]};
dH={Subscript[H,rho][x,y,z],Subscript[H,u][x,y,z],
  Subscript[H,2][x,y,z]};
\end{verbatim}
We then define the bracket we want to consider through the definition of a function
\begin{verbatim}
bracket[a_,b_]:=a.K[b];
\end{verbatim}
where we define the operator
\begin{verbatim}
K[b_]:=Kz[b]-q Kxy[b];
\end{verbatim}
and the corresponding transverse part
\begin{verbatim}
Kxy[b_]:={-bracketXY[psi[[1]],b[[1]]]-bracketXY[
  psi[[2]],b[[2]]]-bracketXY[psi[[3]],b[[3]]],-
  bracketXY[psi[[2]],b[[1]]]-bracketXY[psi[[1]]^3
  psi[[3]],b[[2]]/psi[[1]]]/psi[[1]]+bracketXY[
  psi[[2]],psi[[3]]]b[[3]]/psi[[1]]-bracketXY[
  psi[[1]]^4S3,b[[3]]/psi[[1]]^3]/psi[[1]],-
  bracketXY[psi[[3]],b[[1]]]-bracketXY[psi[[2]],
  psi[[3]]]b[[2]]/psi[[1]]-bracketXY[psi[[1]]^4
  S3,b[[2]]/psi[[1]]]/psi[[1]]^3+3bracketXY[
  psi[[1]]^3psi[[3]]^2,b[[3]]/psi[[1]]^2]/
  psi[[1]]^2-bracketXY[psi[[1]]^5S4,b[[3]]/
  psi[[1]]^3]/psi[[1]]^3-2bracketXY[psi[[2]],
  S3/psi[[1]]^2]b[[3]]-4S3 bracketXY[psi[[2]],
  b[[3]]]/psi[[1]]^2};
\end{verbatim}
and longitudinal part
\begin{verbatim}
Kz[b_]:={-D[b[[2]],z],-D[b[[1]],z]+b[[3]]D[
  psi[[3]],z]/psi[[1]],-b[[2]]D[psi[[3]],z]/
  psi[[1]]-2b[[3]]D[S3,z]/psi[[1]]^2-4S3 D[
  b[[3]]/psi[[1]],z]/psi[[1]]};
\end{verbatim}
The canonical bracket \verb|bracketXY| is defined such that
\begin{verbatim}
bracketXY[a_,b_]:=D[a,x]D[b,y]-D[a,y]D[b,x];
\end{verbatim}
The brackets introduce the closures $S_3(S_2)$ and $S_4(\rho,u,S_2)$ which we define by
\begin{verbatim}
S3=Subscript[S,3][psi[[3]]];
S4=Subscript[S,4][psi[[1]],psi[[2]],psi[[3]]];
\end{verbatim}
It is worth noting that the definition of the function \verb|bracket|, while matching the definition given by Eq.~\eqref{eqBrackS}, does not include the integration. This is done on purpose in order to simplify the code. As shown later, the results will not be affected. \verb|J| is computed by using the command line
\begin{verbatim}
J=bracket[dF,VariationalD[bracket[dG,dH],psi,r]]+
  bracket[dH,VariationalD[bracket[dF,dG],psi,r]]+
  bracket[dG,VariationalD[bracket[dH,dF],psi,r]];
\end{verbatim}
The function \verb|VariationalD[a,b,c]| computes the functional derivative of $\int a(b)\ \mathrm{d}c$, performing operations such as integration by parts by assuming vanishing boundary terms. As the function \verb|VariationalD| acts on the argument of the integral, this justifies the omission of the integral in the definition of the bracket as stated previously. Furthermore, the contributions of second order derivatives of the functionals can be omitted due to the lemma given in Ref.~\cite{Morrison82}.

The Jacobi identity has to be satisfied for any triplet of functionals $(F,G,H)$. We choose in particular $(F,G,H)=(u,\int S_2\ \mathrm{d}\mathbf{r},\int u\ \mathrm{d}\mathbf{r})$. This is done by defining
\begin{verbatim}
dG={0,0,1};
dH={0,1,0};
\end{verbatim}
computing \verb|J| and then computing
\begin{verbatim}
VariationalD[J,dF[[2]],r]
\end{verbatim}
the latter being the translation of $F_\psi=(0,\delta,0)$. However, as the dynamical variables are independent from one another, the terms proportional to $(\partial_y\rho)^2\partial_x^2u$, which are found by using
\begin{verbatim}
CoefficientList[%,{D[psi[[1]],y],
  D[psi[[2]],{x, 2}]}]
\end{verbatim}
have to vanish. This leads to the necessary constraint $\partial S_4/\partial u=0$. Choosing $(F,G,H)=(S_2,\int S_2\ \mathrm{d}\mathbf{r},\int u\ \mathrm{d}\mathbf{r})$ and requiring that the Jacobi identity is satisfied leads to an additional constraint. However, in order to make the term proportional to $(\partial_x\rho)^2\partial_y^2u$ vanish, we have to impose
\begin{equation*}
\rho\frac{\partial S_4}{\partial\rho}+5S_4=4S_3S_3'+9S_2^2,
\end{equation*}
whose solution writes
\begin{equation*}
S_4=\frac{T_4(S_2)}{\rho^5}+\frac{1}{5}\left(4S_3S_3'+9S_2^2\right),
\end{equation*}
where $T_4(S_2)$ is an arbitrary function. We choose $(F,G,H)=(S_2,\int xS_2\ \mathrm{d}\mathbf{r},\int S_2\ \mathrm{d}\mathbf{r})$. In order to make the term proportional to $\partial_x\rho\partial_y\rho\partial_yu$ cancel, we must have
\begin{equation*}
45T_4'+4\rho^5\left(3S_2-(S_3')^2+4S_3S_3''\right)=0,
\end{equation*}
which, as $\rho$ and $S_2$ are independent variables, leads to the necessary constraints $T_4'=0$ and
\begin{equation*}
3S_2-(S_3')^2+4S_3S_3''=0.
\end{equation*}
Computing \verb|J|, one can verify that these constraints are sufficient to ensure that the terms $J_\perp$ and $\{F,\{G,H\}_\perp\}_\parallel+\{F,\{G,H\}_\parallel\}_\perp+\circlearrowleft$ vanish such that Bracket~\eqref{eqBrackS} satisfies the Jacobi identity and hence is of Poisson type.

\bibliography{biblionew}

\end{document}